\documentclass{JHEP3}
\usepackage{epsfig}

\title{ Probing Lepton Flavor Violation Signal Induced by R-violating Minimal
        Supersymmetric Standard Model at a Linear Collider
 \footnote{Supported by National Natural Science Foundation of China.}}
 \vspace{3mm}

\author{{ SUN Yan-Bin$^{b}$, HAN Liang$^{b}$, MA Wen-Gan$^{a,b}$,
    TABBAKH Farshid$^{b}$, ZHANG Ren-You$^{b}$ and ZHOU Ya-Jin$^{b}$}\\
 {\small $^{a}$ CCAST (World Laboratory), P.O.Box 8730, Beijing 100080,
 P.R.China} \\
 {\small $^{b}$ Department of Modern Physics, University of Science and
 Technology}\\
 {\small of China (USTC), Hefei, Anhui 230026, P.R.China}\\
 {\small Email: sunyb@mail.ustc.edu.cn, hanl@ustc.edu.cn}
 }

\abstract{
 The lepton-flavor violation (LFV) effect at an $e^+e^-$ linear
 collider (LC), in the frame of R-parity violating minimal
 supersymmetric standard model is studied. We take the
 R-parity violating processes $e^+e^-\rightarrow e^{\mp}\mu^{\pm}$
 as signal, and define the summation of the two processes
 as ``experiment'' observable. We find that the cross-section summation
 can reach $\cal{O}$$(10^1)fb$ in the parameter space without
 sneutrino resonance effect($\sqrt{s} \sim m_{\tilde{\nu}}$). The
 summation treatment manifests uniform differential distribution
 on $\cos\theta$, where $\theta$ denotes the polar angles of both
 outgoing $e^+/e^-$ respectively to incoming electron beam in two
 signal processes. The uniform feature together with $e\mu$ collinearity
 would help to reduce the SM background dramatically. Consequently we
 conclude that at a $500~GeV$ LC with $480~fb^{-1}$ annual luminosity,
 it's either possible to detect the distinctive R-violating LFV $e\mu$
 signal, or exclude sneutrino to $m_{\tilde{\nu}}>1.1~TeV$ at 95\% CL
 in the machine's biennial runtime interval.}


\begin{document}
\section{Introduction}
\par
Although the Standard Model (SM) is very successful, it's only an effective theory describing physics up to ${\cal
O}(10^2)$GeV. Many experiments have been proposed in the past to find signals of new physics, and many new
theoretical models have been developed to extend physics beyond the SM. Among these extensions of the SM,
supersymmetric models are the most attractive ones, by offering an elegant way to solve the hierarchy problem and
keep a consistent unification of gauge couplings.
\par
In the SM the conservations of the baryon number $B$ and lepton number $L$ are automatic consequences of the gauge
invariance and renormalizability. On the other hand, the lepton number conservation for individual generation has no
strong theoretical basis. In addition, recent neutrino oscillation experiments \cite{experi1,experi2,experi3}
manifest that neutrinos strongly mix among flavors. Therefore, to accommodate the observation of neutrino
oscillation which is apparently lepton flavor violating (LFV), the SM has to be extended. As the most general
minimal supersymmetric extension of the SM, the R-parity violating minimal supersymmetric standard model
($\rlap/R_{p}$-MSSM) contains all renormalizable terms which observe the $SU(3)_C\otimes SU(2)_W \otimes U(1)_Y$
gauge-symmetry, supersymmetry, and the most general superpotential with R-parity violation given by
\cite{potential1}
\begin{equation}
\label{sup}
{\cal W}_{\rlap/R_{p}} =
\frac{1}{2}\epsilon_{ab} \lambda_{ijk}\hat{L}_{i}^a \hat{L}_{j}^b \hat{E}_{k} +
\epsilon_{ab}\lambda^{'}_{ijk} \hat{L}_{i}^a \hat{Q}_{j}^b \hat{D}_{k} +
\frac{1}{2}\epsilon_{\alpha\beta\gamma}\lambda^{''}_{ijk}
   \hat{U}_{i}^{\alpha} \hat{D}_{j}^{\beta} \hat{D}_{k}^{\gamma} +
\epsilon_{ab}\delta_{i} \hat{L}_{i}^a \hat{H}_{2}^b
\end{equation}
where $i,j,k=1,2,3$ are generation indices, $a,b=1,2$ are SU(2)
isospin indices, and $\alpha,\beta,\gamma=1,2,3$ are SU(3) color
indices. $\lambda,\lambda^{'},\lambda^{''}$ are dimensionless
R-violating Yukawa couplings with
$\lambda_{ijk}=-\lambda_{jik}$, $\lambda_{ijk}^{\prime
\prime}=-\lambda_{ikj}^{\prime \prime}$. The last bilinear terms
mix the lepton and the Higgs superfield which may generate masses of
neutrinos.

\par
Since the first two terms, $\hat{L}\hat{L}\hat{E}$ and $\hat{L}\hat{Q}\hat{D}$, in the R-parity violating
superpotential may lead to single sneutrino production and sequential LFV final states, they are of special
interest. Many authors studied these sneutrino s-channel production modes, both on- and off-mass-shell of sneutrino
resonance \cite{Rstrain,snuTevatron,snuLHC,snuLCbiterm}. Most of these works focused on how to probe the resonance
effect on hadron colliders, such as Tevatron and LHC, where sneutrinos are produced via $\hat{L}\hat{Q}\hat{D}$
interactions and decay in R-parity conservation mode with high branch ratio, such as $\tilde{\nu} \rightarrow
l\tilde{\chi}^\pm_i$. The signal of these sneutrino R-parity violating production and subsequential R-conserving
decay modes, includes three leptons in final state, where two of the leptons are from chargino/neutrolino cascade
decay. Obviously the analysis of these tri-lepton events is a good way to discover new physics beyond the SM.
However, tri-lepton production can also be induced by some other new physics models, for example even in
R-conserving MSSM, $\tilde{\chi}^\pm_i \tilde{\chi}^0_j$ association production may have the subsequential decay to
three leptons plus two $\tilde{\chi}^0_1$ as the lightest supersymmetric particle (LSP). So, tri-lepton signal might
not be able to distinguish R-violating interaction from other `new physics background'.

\par
In this paper, the possibility of detecting the di-lepton LFV
processes $e^+e^-\rightarrow e^\mp\mu^\pm$ at an electron-positron
linear collider(LC) is discussed in the framework of the minimal
supersymmetric standard model(MSSM) with R-parity violation. We
will demonstrate that with clean collision environment and high
luminosity, an energetic $e^+e^-$ LC machine will be a powerful
tool to discover the R-parity violating LFV interactions.

\vskip 5mm
\section{The calculation of the processes $e^+e^-\rightarrow
  e^\mp\mu^\pm$ in R-violating MSSM}
\par
The R-violating interactions relevant to the calculations of the
processes $e^+e^-\rightarrow e^\mp\mu^\pm$ at the leading order,
are given by $\hat{L}\hat{L}\hat{E}$-type terms of the
superpotential. By integrating Eq.(\ref{sup}) over
supercoordinates($\theta$, $\bar{\theta}$), one obtains the
tri-lepton lagrangian
\begin{equation}
\label{lag}
{\cal L}_{\hat{L}\hat{L}\hat{E}}=
\lambda_{ijk}\cdot (\bar \nu_{i}^c P_L l_{j}\tilde l_{R k}^* +
\bar l_{k} P_L \nu_{i}\tilde l_{L j}+
\bar l_{k} P_L l_{j}\tilde \nu_{L i})~+~h.c.
\end{equation}
where $P_{L/R}=(1\mp\gamma_5)/2$ are left/right-hand
project operator, and $c$ refers to charge conjugation.

\par
\begin{figure}[htb]
\centering
\epsfig{file=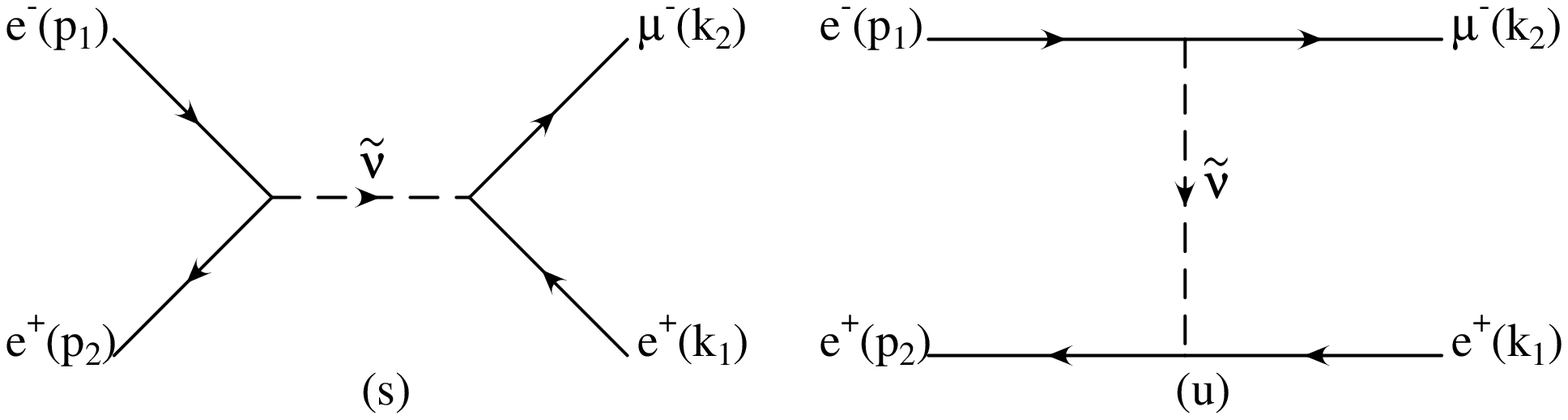,width=280pt,height=80pt}
\epsfig{file=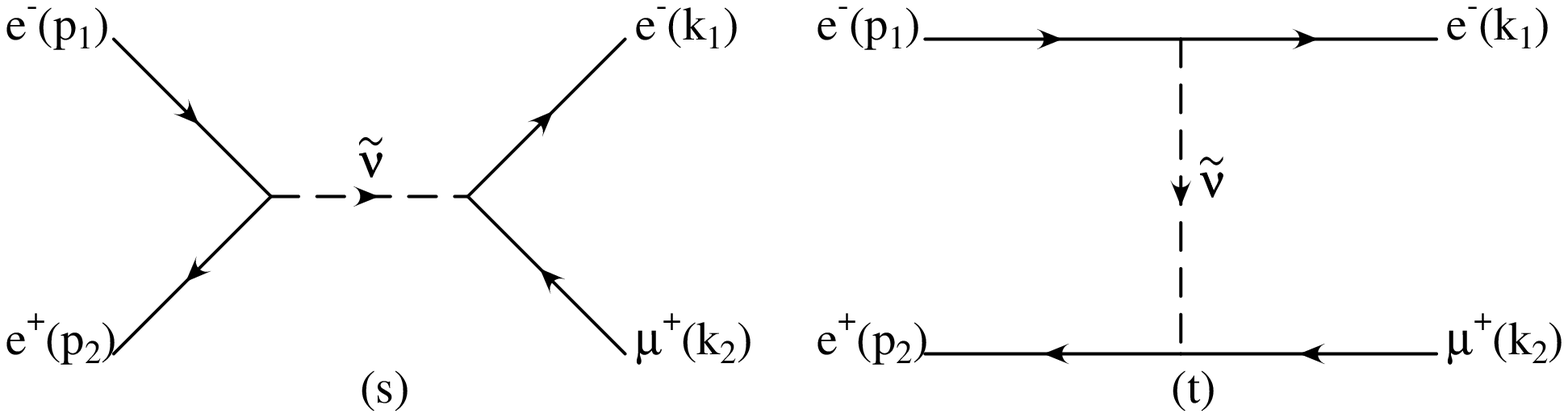,width=280pt,height=80pt}
\vspace*{-10pt} \caption{\scriptsize The relevant Feynman diagrams
to the processes of $e^+e^-\rightarrow e^+\mu^-$ and
$e^+e^-\rightarrow e^-\mu^+$. The upper two diagrams are
corresponding to the $e^+ \mu^-$ production, and the lower two
are to the $e^- \mu^+$ production. With the definition of
$\theta$, $e^+\mu^-$ process is represented as "s" and "u"
channel, while $e^-\mu^+$ as "s" and "t" channel.} \label{fFeym}
\end{figure}

The Feynman diagrams of LFV signal processes
$e^+e^-\rightarrow e^\mp\mu^\pm$
are depicted in Fig.\ref{fFeym}\footnote{Here we do not present
the diagrams which can be obtained by reversing the current arrow
of $\tilde{\nu}$}.
From the experimental point of view, the charge measurement of
high transverse momentum ($p^T$) tracks relies on many realistic
factors, such as the configuration of the magnetic field
surrounding tracker subdetector, the radius of outmost tracker
system in transverse plane, and the spatial resolution of the
tracker etc. The higher the track $p_T$ is, the less accurate the
lepton charge is determined. In order to get optimal efficiency on
signal detection, we deliberately do not choose the criteria of
using opposite charge veto in event selection, i.e. di-tracks of
$e^-\mu^+$ final state won't be distinguished from those of
$e^+\mu^-$. To reflect this `non-signed' $e\mu$ experiment
measurement, we use a consistent momentum notation to denote the
two signal processes as:
\begin{equation}
\label{process1} e^-(p_1) + e^+(p_2) \rightarrow
e^-(k_1)+\mu^+(k_2),
\end{equation}
\begin{equation}
\label{process2}e^-(p_1) + e^+(p_2) \rightarrow e^+(k_1) +
\mu^-(k_2),
\end{equation}
where $p_{1}$ and $p_{2}$ are four-momenta of incoming electron
and positron beams, $k_1$ represents outgoing $e^-(e^+)$ momenta
in the two signal processes, and $k_2$ denotes that of
$\mu^+(\mu^-)$ final particle correspondingly. The Mandelstam
variables are defined accordingly, where electron and muon are
taken as massless for simplicity.
\begin{eqnarray}
 s &=& (p_1 + p_2)^2=(k_1 + k_2)^2\nonumber\\
 t &=& (p_1 - k_1)^2=(p_2 - k_2)^2 = -\frac{s}{2}(1-\cos\theta)\\
 u &=& (p_1 - k_2)^2=(p_2 - k_1)^2 = -\frac{s}{2}(1+\cos\theta)\nonumber
\end{eqnarray}
where $\theta$ is defined to denote the polar angle of outgoing
$e^-$($e^+$) of the individual signal process $e^+e^-\rightarrow
e^-\mu^+$ ($e^+e^-\rightarrow e^+\mu^-$) with respect to the
incoming electron beam. We sum up the production events of the two
signal processes given in Eqs.(\ref{process1}-\ref{process2}),
which have equal cross sections and asymmetric differential ones
due to CP-conservation and the above $\theta$ definition. This
summation treatment of the two signal processes is consistent with
the `non-signed' $e\mu$ experiment observation, where the charges
of the final $e$ and $\mu$ particles are not distinguished.

\par
In this paper we assume that the sneutrino mass spectrum is degenerate, i.e.
\begin{equation}
m_{\tilde\nu_i} = m_{\tilde\nu} ~~~~~~(i=1,2,3)
\end{equation}
The amplitude of the $e^-\mu^+$ production process is denoted as
\begin{equation}
\label{amp0}
{\cal M}^{(-)} = \sum_{i=2,3} ( {\cal M}^{(-)}_{s,i} - {\cal
M}^{(-)}_{t,i} )
\end{equation}
with
\begin{eqnarray}
\label{amp1} {\cal M}^{(-)}_{s,i} &=&
  -i~\lambda_{i11}\lambda_{i12}\cdot
  \bar v(p_2) P_L u(p_1) \cdot {\cal P}(s,m_{\tilde\nu}) \cdot
   \bar u(k_1)P_R v(k_2)\nonumber\\
&&+~(P_L \leftrightarrow P_R,~\lambda_{i12}\rightarrow\lambda_{i21})\nonumber\\
{\cal M}^{(-)}_{t,i} &=&
  -i~\lambda_{i11} \lambda_{i12}\cdot
   \bar v(p_2)P_R v(k_2) \cdot {\cal P}(t,m_{\tilde\nu}) \cdot
    \bar u(k_1)P_L u(p_1)\nonumber \\
&&+ ~(P_L \leftrightarrow P_R,~\lambda_{i12}\rightarrow
\lambda_{i21}).
\end{eqnarray}
Analogically, the amplitude of the $e^+\mu^-$ production process is given as
\begin{equation}
{\cal M}^{(+)} = \sum_{i=2,3}({\cal M}^{(+)}_{s,i} - {\cal
M}^{(+)}_{u,i}),
\end{equation}
with
\begin{eqnarray}
\label{amp2}
{\cal M}^{(+)}_{s,i} &=&
 -i~\lambda_{i11} \lambda_{i21}\cdot \bar v(p_2)P_L u(p_1) \cdot
  {\cal P}(s,m_{\tilde{\nu_i}}) \cdot \bar u(k_2)P_R v(k_1)\nonumber\\
&&+~(P_L \leftrightarrow P_R,~\lambda_{i21}\rightarrow \lambda_{i12})
  \nonumber\\
{\cal M}^{(+)}_{u,i} &=&
 -i~\lambda_{i12} \lambda_{i11}\bar v(p_2) P_R v(k_1) \cdot
  {\cal P}(u,m_{\tilde{\nu_i}}) \cdot \bar u(k_2)P_L u(p_1)\nonumber \\
&&+~(P_L \leftrightarrow P_R,~\lambda_{i12}\rightarrow
\lambda_{i21}).
\end{eqnarray}
Here, due to the coupling strength of sneutrino to the incoming electron and positron beam, only the generation
index $i=2,3$ of the propagating sneutrino are considered. Under mass degeneration assumption, propagators of
sneutrinos $\tilde\nu_{2,3}$ are given as
\begin{eqnarray}
{\cal P}(t,m_{\tilde\nu}) &=& \frac{1}{t -
m_{\tilde\nu}^2} \nonumber\\
{\cal P}(u,m_{\tilde\nu}) &=& \frac{1}{u -
m_{\tilde\nu}^2} \nonumber\\
\label{propg} {\cal P}(s,m_{\tilde\nu}) &=& \frac{1}{s -
m_{\tilde\nu}^2+i s \Gamma_{\tilde\nu}/m_{\tilde\nu}}
\end{eqnarray}
Here, to suppress resonance enhancement the sneutrino decay width
$\Gamma_{\tilde\nu}$ is taken into account in the s-channel
propagator .

\vskip2mm
\par
Using Eqs.(\ref{amp0}-\ref{propg}), the summation of the differential cross sections of the two processes $e^+e^-\to
e^-\mu^+$ and $e^+e^-\to e^+\mu^-$ can be written as
\begin{eqnarray}
\label{dsig} \frac{d\sigma_{e\mu}}{d\cos\theta} &=&
 \frac{1}{4}\frac{1}{32 \pi s} \{ |
  {\cal M}^{(-)} |^2 + | {\cal M}^{(+)}|^2 \}\nonumber\\
&=& \frac{1}{4}\frac{1}{32 \pi s}
(\lambda_{211}^2\lambda_{212}^2 +
 \lambda_{311}^2\lambda_{312}^2 +
 \lambda_{311}^2\lambda_{321}^2 +
2\lambda_{211}\lambda_{212}\lambda_{311}\lambda_{312})\nonumber\\
&~~& \times (2s^2\cdot |{\cal P}(s,m_{\tilde\nu})|^2 +
             t^2 \cdot |{\cal P}(t,m_{\tilde\nu})|^2 +
             u^2\cdot |{\cal P}(u,m_{\tilde\nu})|^2)
\end{eqnarray}
where the subscript $e\mu$ denotes the ($e^-\mu^+) + (e^+\mu^-$)
signal summation treatment. Due to the R-violating
scalar-pseudoscalar(S-P) Yukawa couplings, the interference
contribution among different topological diagrams vanishes as
shown in the above equation. Another thing to be mentioned here is
that the summation treatment of the $e^\pm\mu^\mp$ production
processes doubles the 's'-channel contributions, while
't+u'-channel contributions cancel their individual
forward-backward asymmetry on $\cos\theta$.

\vskip 5mm
\section{Numerical results and discussion}
\subsection{Signal}
\par
Required by superpotential Eq.(\ref{sup}), $\lambda_{iij}$ should be zero. For the other R-parity violating
parameters, we refer to the experimental constraints presented in Ref.\cite{Rstrain}, and take the values of the
$\lambda_{ijk}$ coupling parameters as
\begin{eqnarray}
\lambda_{12j}=-\lambda_{21j}=0.049~,&&
  \lambda_{31j}=-\lambda_{13j}=0.062 \nonumber\\
\lambda_{23j}=-\lambda_{32j}=0.070~,
 \label{rparam}
\end{eqnarray}
It should be stressed that these $\lambda$ parameters are SUSY mass
dependent, e.g. they could be scaled by a factor of
$m_{\tilde{l}_jR}/100$[GeV]. In this work, a degenerate charged
slepton spectrum of $100$GeV is assumed, which could keep R-violating
Yukawa couplings less than $\cal{O}$$(10^{-1})$.
\par
To make our scenario consistent with available $e^+e^-$ collision
data, the degenerated sneutrino mass has to be constrained. For
example, OPAL Experiment has set an upper limit on
$\sigma(e^+e^-\rightarrow e \mu)$ as $22~fb$ with $200\le\sqrt{s}
\le 209~GeV$ at 95\% CL \cite{OPAL}. To get a comparable value in
this energy range by applying Eq.(\ref{dsig}) with the input
R-violating parameters given in Eq.(\ref{rparam}), one can arrive
at
$$
m_{\tilde\nu} \ge 250\mbox{GeV}.
$$

%\vskip2mm
\par
\begin{figure}[htb]
\centering
\epsfig{file=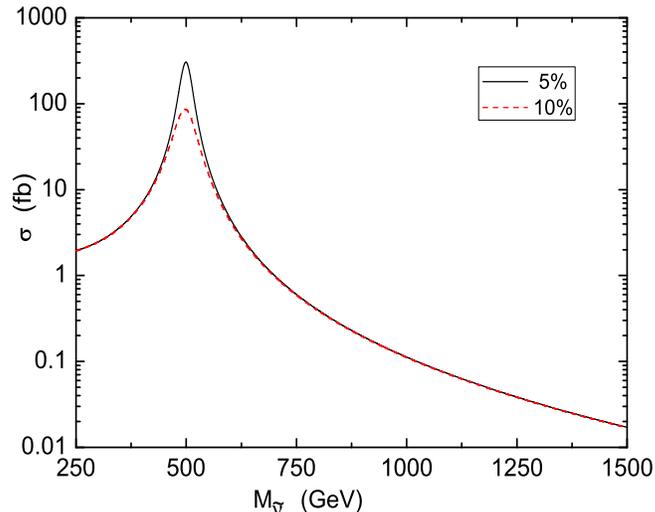,width=280pt,height=240pt}
\vspace*{-30pt} \caption{\footnotesize LFV cross section $\sigma_{e\mu}$
as function of $m_{\tilde{\nu}}$ at $\sqrt{s}=500$GeV. Solid and
dashed line correspond to $\Gamma_{\tilde{\nu}}=5,10\%~
m_{\tilde{\nu}}$ respectively.}
\label{fSigem}
\end{figure}

Since linear colliders are running at fixed c.m.s energy, one can't expect to be so lucky that $\sqrt{s}$ will be so
close to the sneutrino mass that real sneutrinos are produced or large resonance enhancement will occur. Therefore,
in this work we mainly discuss the range of the sneutrino mass window that can be probed via off-resonance $e\mu$
LFV signal, at a LC collider running at `moderate' energy such as TESLA Run1. Setting $\Gamma_{\tilde{\nu}}=5\%$ and
10\% of sneutrino mass $m_{\tilde{\nu}}$ to block on-resonance effect, we calculate the dependence of
$\sigma_{e\mu}$ on $m_{\tilde{\nu}}$ at $\sqrt{s}=500~GeV$, and depict it in Fig.\ref{fSigem}.

\par
From Fig.\ref{fSigem} one can see that at a $500~GeV$ LC machine,
even without resonance enhancement the R-violating LFV
cross-section of signal summation may reach $\cal{O}$(1)$fb$,
which could be detected clearly at high luminosity linear
colliders. However, if degenerated sneutrinos are heavier than
$1~TeV$, the signal will decrease to no more than 0.1$fb$, which
might be overwhelmed by background. Now, the question is whether
some event selection strategies can be developed to suppress
background efficiently, so that off-resonance sneutrino LFV effect
could be detected and sneutrino mass parameter limit would be
extended up to $\cal{O}$(TeV). In the following discussion we will
focus on the signal of $m_{\tilde{\nu}}=1~TeV$ contribution.

\par
\begin{figure}[htb]
\centering
\epsfig{file=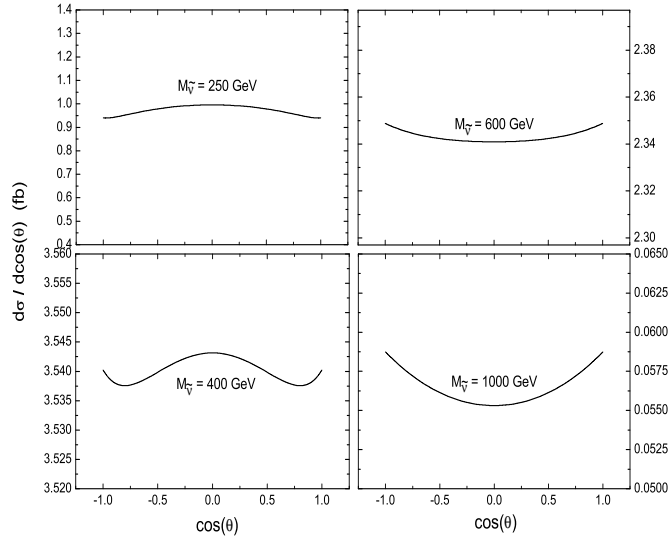,width=280pt,height=240pt}
\vspace*{-30pt} \caption{\footnotesize LFV differential cross section
$\frac{d\sigma_{e\mu}}{d\cos\theta}$ at $\sqrt{s}=500$GeV for
different sneutrino masses.} \label{fdSigem}
\end{figure}

\par
\begin{figure}[htb]
\centering
\epsfig{file=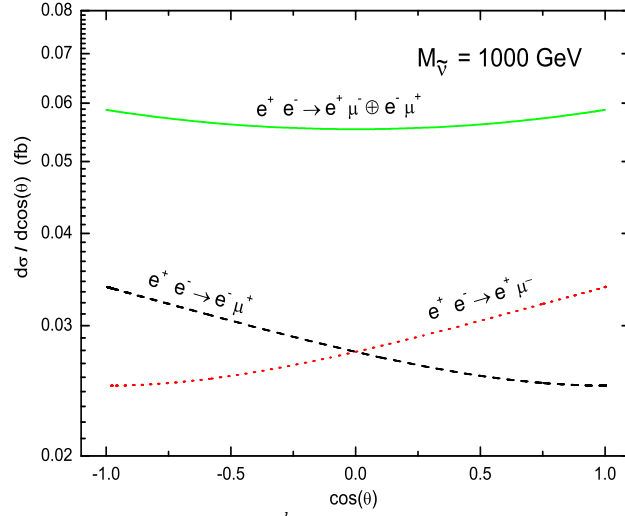,width=280pt,height=240pt}
\vspace*{-30pt} \caption{\footnotesize LFV differential cross section
$\frac{d\sigma_{e\mu}}{d\cos\theta}$ at $\sqrt{s}=500$GeV
contributed by 1TeV sneutrinos.} \label{fdSigem1T}
\end{figure}

The angular distributions of four typical $m_{\tilde{\nu}}$ values are given in Fig.\ref{fdSigem}. The differential
cross sections are nearly uniform versus $\cos\theta$ merely with trivial fluctuation. This uniformly distributing
feature is mainly derived by the $e\mu$ summation treatment, namely though the two signal processes $e^+e^- \to
e^+\mu^-$ and $e^+e^- \to e^-\mu^+$ tend to forward and backward respectively, the summation would erase the
individual tendency. The cancellation determined by the summation treatment manifests itself clearly in
Fig.\ref{fdSigem1T}, where the differential cross sections of both processes $e^+e^- \to e^+\mu^-$, $e^+e^- \to
e^-\mu^+$ and the summation contribution at $m_{\tilde{\nu}}=1~TeV$ are plotted in the same frame.

\par
\begin{figure}[htb]
\centering
\epsfig{file=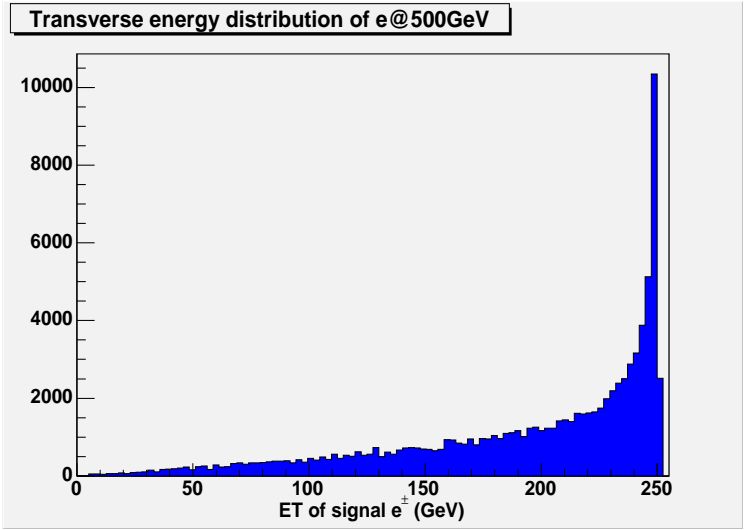,width=280pt,height=200pt}
\vspace*{-10pt} \caption{\footnotesize Transverse energy
distribution of LFV signal of $m_{\tilde{\nu}}$=1TeV at
$\sqrt{s}=500$GeV. Y-axis is arbitrary scale} \label{fSigpT}
\end{figure}

Correspondingly, Fig.\ref{fSigpT} shows the transverse momentum
distribution of the signal. It is clear that the outgoing $e$ and
$\mu$ transverse momenta will rush to the beam energy $E_{beam}$,
which is consistent with the uniform distribution of the
differential cross section $\frac{d\sigma_{e\mu}}{d\cos\theta}$.
It will be demonstrated below that the uniform distribution
feature versus $\cos\theta$ of the LFV differential cross-section
is very useful for extracting the signal from the background.

\subsection{SM background and Event selection}
In the numerical calculation of physical background, the SM input
parameters are taken as $m_{\tau}=1776.99~MeV$, $m_W=80.423~GeV$,
$m_Z=91.1876~GeV$ and $\alpha=1/128$\cite{SMpa}.

\par
\begin{figure}[htb]
\centering
\epsfig{file=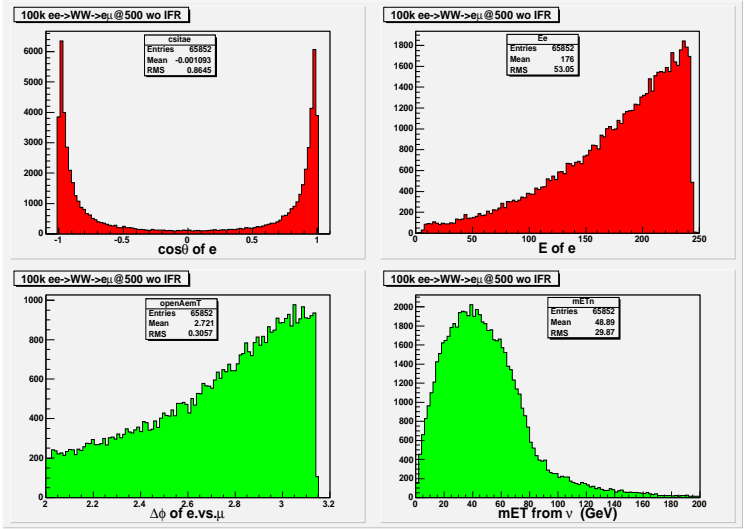,width=280pt,height=200pt}
\vspace*{-10pt} \caption{\footnotesize MC distributions of $e(\mu)$ from
SM background $e^+e^-\rightarrow WW \rightarrow e\mu+X$ at
$\sqrt{s}=500~GeV$. $\cos\theta$ and energy of final $e$, $\phi$
separation of $e\mu$ and missing transverse energy are plotted}
\label{fMCWW}
\end{figure}
\begin{figure}[htb]
\centering
\epsfig{file=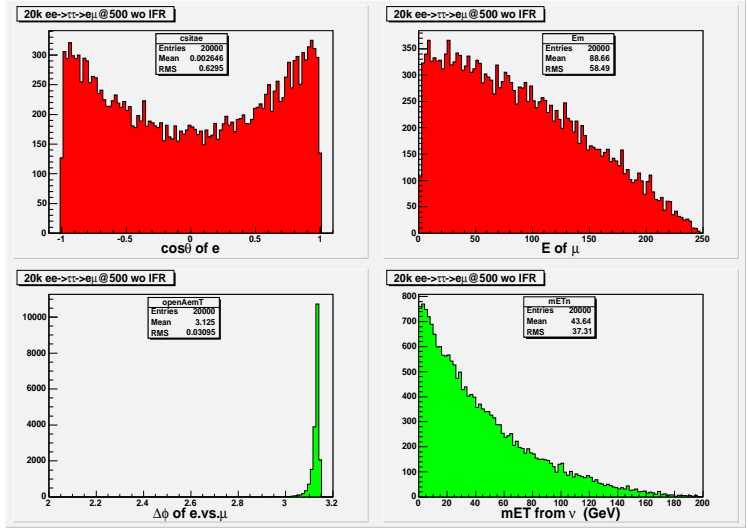,width=280pt,height=200pt}
\vspace*{-10pt} \caption{\footnotesize MC distributions of $e(\mu)$ from
SM background $e^+e^-\rightarrow \tau\tau \rightarrow e\mu+X$ at
$\sqrt{s}=500~GeV$.} \label{fMCTT}
\end{figure}

Due to its clean collision environment, LC is considered as
powerful facility for precise studies on particle physics, which
will complement and extend the physics program of hadron colliders
such as Tevatron and LHC. One can expect that there will be an
ideally good detector on LC, especially high performance central
calorimeter and muon-tracker system with large angular coverage
and good energy/transverse-momentum resolution. In such a capable
detector, QCD instrumental background where two hadronic jets fake
both energetic electron and muon candidate in same event, is
negligible. Therefore, the main background to the LFV signal is
from physical processes $$ e^+e^-\rightarrow W^+W^-,
\tau\bar{\tau},b\bar{b},t\bar{t} \rightarrow e\mu+X $$ where $X$
refers to decay products of $W,\tau,b$ and $t$ other than $e\mu$.
Among them, the last two ones can be removed easily: since the
$e\mu$ from $b\bar{b}$ are very soft and always associated with
jets as $b \rightarrow q W^{*} \rightarrow jl\nu$, one is able to
eliminate this kind of background by some moderate
calorimeter-based energy isolation cut on both $e\mu$ candidates.
For $t\bar{t}$ events, they can be rejected by veto on high energy
jets from two b-quarks. So, only $WW$ and $\tau\bar{\tau}$
background have to be taken into account here. In this section, an
event selection strategy will be introduced to reduce these two
sorts of background, and its efficiency on the background is
compared with that on the signal. The $WW$ contribution at
$\sqrt{s}=500~GeV$ is given by
\begin{eqnarray}
\sigma_{WW} &=&
 \sigma[e^+e^-\rightarrow W^+W^-]\times 2 \cdot
  \mbox{Br}[W\rightarrow e\nu_e] \cdot
  \mbox{Br}[W\rightarrow \mu\nu_{\mu}] \nonumber \\
 ~ &\sim& {\bf 162.5} ~fb \nonumber
\end{eqnarray}
Some MC distributions of $WW\rightarrow e\mu$ background generated
by Pythia\cite{pythia}, are plotted in Fig.\ref{fMCWW}. The
prominent feature of the WW background is the strong
$\cos\theta$ dependence.
The background from $\tau\tau$ is around ${\bf 28}~fb$, and MC
distributions are plotted in Fig.\ref{fMCTT}. The final $e\mu$
from di-$\tau$ are much softer than those from WW and
signal as well.

\vskip2mm
\par
An "off-line" three-step event selection strategy is invented as follow:
%\begin{itemize}
\begin{enumerate}

\item
$e\mu$ from $WW$ production are almost forward-backward distributed, while those from signal incline perpendicular
to the beam. So we define {\bf CUT1} on polar angles as
\begin{equation}
0.55~\le~\theta_l~\le~(\pi-0.55)~~~~l=e,\mu
\end{equation}

\item
To have sensitivity to events produced at effective beam energy lower than
the actual $\sqrt{s}/2$, we accept events which pass following loose {\bf CUT2}
on energy and momentum
\begin{eqnarray}
E_e&\ge&85\%\cdot E_{beam},\nonumber\\
p_{\mu}&\ge&75\%\cdot E_{beam}
\end{eqnarray}
where for the electron the energy $E_e$ is measured in
calorimeter, and for the muon the momentum $p_{\mu}$ is determined
by central tracker system within a magnetic field.

\item
Assuming high spatial resolution on x-y plane vertical to the beam, we
set a severe {\bf CUT3} on $e\mu$ $\phi$-difference to select
back-to-back events as
\begin{equation}
|\pi-\Delta\phi_{e\mu}|~\le~0.01
\end{equation}

%\end{itemize}
\end{enumerate}
It's apparent that this three-step CUT strategy takes advantage of the uniform
differential distribution and collinearity feature of LFV signal.
Here considering relatively poor muon track $p_{\mu}^T$
resolution, we won't use missing transverse energy cut.

\par
\vskip2mm Three sets of MC samples for $WW$, $\tau\tau$ and signal
productions via $e^+e^-$ collision at $\sqrt{s}=500~GeV$ are
generated by Pythia with different assumed luminosities. Each
process is simulated with the three-step event selection, and the
numbers of events passing individual CUT are listed in Table \ref{TableSB},
respectively.
\begin{table}[htb]
\centering
\begin{tabular}{|l|c|c|c|}
\hline
 &SM  background & SM background & $e-\mu$ signal\\
& $WW\rightarrow e\mu$ & $\tau\tau\rightarrow e\mu$ &($m_{\tilde{\nu}}=1$TeV)\\
\hline\hline
no cut $N_0$ & 65852 & 20000 &90777 \\
(at L $fb^{-1}$) & (4E2) & (7E2) & (8E5) \\
\hline
{\bf CUT1} $N_1$ & 15749 & 15864 &76269 \\
\hline
{\bf CUT2} $N_2$ & 2131 & 28 & 76269 \\
\hline
{\bf CUT3} $N_3$ & 118  & 28 & 76269 \\
\hline
efficiency $\epsilon$ & 0.179\% & 0.14\% & 84.0\% \\
\hline\hline
$\sigma$ before CUT  & 162.5 $fb$ & 28.1 $fb$ & 0.113 $fb$ \\
\hline
$\sigma$ after CUT   & 0.291 $fb$ & 0.039$fb$ & 0.095 $fb$ \\
\hline
\end{tabular}
\caption{\footnotesize Event selection efficiency on background and
signal. The first 4 rows are numbers of events before and after
individual CUT. The values of event selection efficiency on different
samples are given in the 5th row.}
\label{TableSB}
\end{table}
%\vskip1mm

\par
Despite different integrated luminosities in generating
the MC samples, we can define an unitary event selection
efficiency on both background and signal as
\begin{equation}
\epsilon= \frac{N_3}{N_0}
\end{equation}
It's demonstrated that with the three-step strategy we are able to reduce the background by 3 orders, i.e.
$\epsilon_{WW} \sim 0.18\%$ and $\epsilon_{\tau\tau} \sim 0.14\%$ , while keeping the selection efficiency on the
signal as high as $84\%$. The selection would result in an approximate $0.1 fb$ signal cross section of 1$TeV$
sneutrino, which is about 3 factors smaller than the SM background. Since these CUTs have already left room for real
beam at LCs and detector performance and are unbiased to both signal and background, we assume that the cross
sections after CUTs are close to full efficiency with respect to a given luminosity.

\par
\begin{figure}[htb]
\centering
\epsfig{file=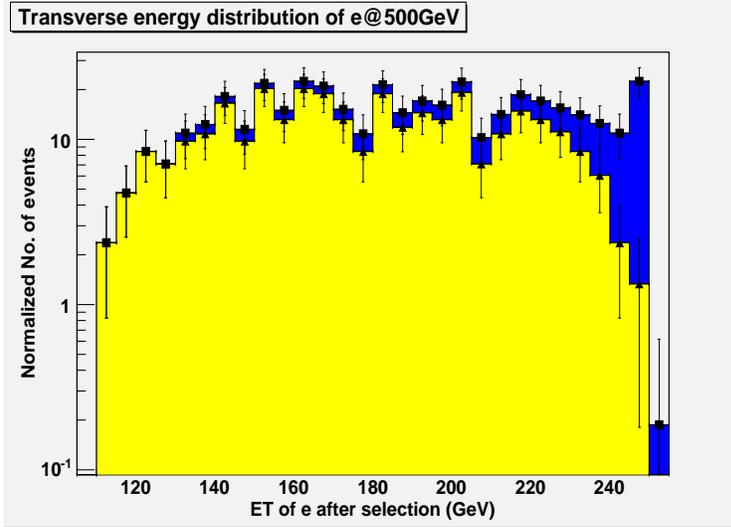,width=280pt,height=200pt}
\vspace*{-10pt} \caption{\footnotesize MC ET distributions of $e$
candidates after event selection at $\sqrt{s}=500$GeV. The number
of events has been normalized by 960$fb^{-1}$ luminosity. YELLOW
is for $WW$ and $\tau\tau$ SM background; BLUE is for background
plus $\rlap/ R$-violating signal with $m_{\tilde{\nu}}=1~TeV$.}
\label{fCUTleET}
\end{figure}
\begin{figure}[htb]
\centering \epsfig{file=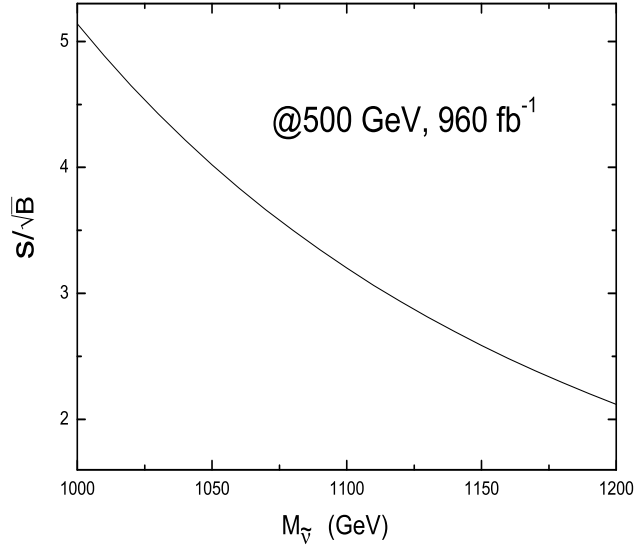,width=280pt,height=260pt}
\vspace*{-30pt} \caption{\footnotesize The dependence of
R-violating signal significance on sneutrino masses.} \label{fSB}
\end{figure}

At a 500 GeV $e^+e^-$ collider with certain integrated luminosity L, the number of background events B and that of
signal S contributed by sneutrinos with $m_{\tilde{\nu}}=1~TeV$ after selection are given by
\begin{eqnarray}
\mbox{S} &=& (\sigma_{e\mu} \cdot \epsilon_{e\mu}) \cdot L
         = \sigma_{e\mu}^{CUT} \cdot L \nonumber\\
     &=&  0.095~fb \cdot L
\end{eqnarray}
\begin{eqnarray}
\mbox{B} &=& (\sigma_{WW} \cdot \epsilon_{WW} +
            \sigma_{\tau\tau} \cdot \epsilon_{\tau\tau}) \cdot L
        = \sigma_{SM}^{CUT} \cdot L
        \nonumber\\
        &=& 0.33~fb \cdot L
\end{eqnarray}
where $\sigma_{SM}^{CUT}$ is the sum of $WW$ and $\tau\tau$ contributions after CUTs, and $\sigma_{e\mu}^{CUT}$ is
the cross-section of R-violating LFV signal respectively. The value of S/B is about 0.3 which is acceptable for a
stable signal/background analysis. Sequentially, the significance of signal over background is defined as
\begin{equation}
\mbox{SB} = \frac{S}{\sqrt{B}}
       = \frac{\sigma_{e\mu}^{CUT}}{\sqrt{\sigma_{SM}^{CUT}}}
       \cdot \sqrt{L}
\end{equation}
Supposing a typical luminosity at a LC can reach $2\times 10^{34}
cm^{-2} s^{-1}$\cite{lumi}, it's reasonable to presume a
$480~fb^{-1}$ annual data acquisition. Thereby, the biennial data
accumulation on LC that is greater than $940~fb^{-1}$, will
provide a significance as large as 5 which is sufficient for the
discovery of the R-violating LFV interaction. The transverse
energy($E^T$) distribution of electron candidates accumulated per
year is simulated in Fig.\ref{fCUTleET}. The high $E^T$ tendency
of the signal is obvious and distinctive from SM background.
The significance varying with different sneutrino mass values at
$\sqrt{s}=500$GeV is given in Fig.\ref{fSB}, which decreases with
the increment of sneutrino mass and would drop to 2.5 at
$m_{\tilde{\nu}}\sim 1.15~TeV$, namely if no clue of signal is
seen one can exclude sneutrinos up to this mass scale at 95\% CL.
\par
Finally, we want to say some words on potential backgrounds contributed by other non-SM physics. Even without
R-parity violation, MSSM can induce large di-lepton LFV effect at LC collider\cite{LFV1slmix,LFV2sloop} too. Typical
processes are heavily-mixed slepton pair production and cascade decay, $e^+e^-\rightarrow \tilde{l}^+_i
\tilde{l}^-_i \rightarrow e\tilde{\chi}^0_1 +\mu\tilde{\chi}^0_1$. The character of these background is moderate $e$
and $\mu$ associated with large missing $E^T$ carried by LSP $\tilde{\chi}^0_1$, and these R-conservation LFV
background can be removed from the $\rlap/{R}_p$ event samples by $e\mu$ energy CUT2 and colinearity CUT3. Some
other interesting non-SM models predict a heavy $Z^{'}$ which could couple to $\mu\tau$, $e\tau$ or even $e\mu$
\cite{LFV3ZpVA}. Due to V-A couplings of $Z^{'}$, the final $e$ and $\mu$ from these background should be severely
forward-backward, and would be cut off by the $\cos\theta$ CUT1. Even if a few background events survive the
three-step off-line LFV selection, since the uniform differential distribution of the $\rlap/{R}_p$-MSSM LFV will
drive transverse momenta of outgoing $e$ and $\mu$ peak at high energy very close to $E_{beam}$ as shown in
Fig.\ref{fCUTleET}, this distinctive feature of R-violating signals will help to estimate the $Z^{'}$ contribution
in the total selected samples.

\section{Summary}
We have studied the lepton flavor violating processes $e^+e^-\rightarrow e^{\mp}\mu^{\pm}$ in the MSSM with R-parity
violation at an electron-positron LC with 2$\times 250~GeV$ colliding energy. To be consistent with experimental
measurement, we sum the two signal processes up and define an observable $\theta$ to denote both final $e^-$ and
$e^+$ polar angles with respect to the eletron beam. The summation cross-section measurement can reach
$\cal{O}$$(10^1)fb$ without apparent sneutrino resonance enhancement.
\par
Determined by scalar-pseudoscalar Yukawa couplings of sneutrino to
leptons, and then enlarged by our $e^{\pm}\mu^{\mp}$ summation
treatment, the R-violating LFV signal is characterized with a
uniform differential distribution onto $\cos\theta$. Using this
uniform distribution feature together with collinearity of $e\mu$
final states, we develop a three-step event selection strategy.
Under this strategy, the SM background can be under control.
Consequently, at a $500~GeV$ LC machine running with annual
luminosity $L_Y=480~fb^{-1}$, one can expect to extend R-violating
interaction search to heavy off-mass-shell sneutrino contribution,
namely in LC's two-year-run either detect $e\mu$ signal induced by
sneutrino with $m_{\tilde{\nu}}=1.0~TeV$ at 99.99\% CL discovery
level, or exclude sneutrino to $m_{\tilde{\nu}}>1.1~TeV$ at $95\%$
CL.

\vskip 3mm \noindent{\large {\bf Acknowledgement:}} This work was
supported in part by the National Natural Science Foundation of
China. The authors would like to thank Prof. Franz F.
Sch$\ddot{o}$berl, University of Wien, for the useful discussion
and comment.

\end{document}